\documentclass{elsart}

\usepackage{times}

 \usepackage{graphics}
\usepackage{graphicx}
\usepackage{epsfig}

\usepackage{amssymb}
\usepackage{amsbsy}
\begin{document}

\begin{frontmatter}


\title{The Crossover between Liquid and Solid Electron Phases in Quantum Dots:
A Large-Scale Configuration-Interaction Study}

\author[ia1]{Massimo Rontani\corauthref{cor1}}
\corauth[cor1]{Corresponding Author:}
\ead{rontani@unimore.it}
\author[ia1,ia2]{Carlo Cavazzoni}
\author[ia1,ia3]{Guido Goldoni}
\address[ia1]{INFM National Research Center S3,
Modena, Italy}
\address[ia2]{CINECA, Casalecchio di Reno, Italy}
\address[ia3]{Dipartimento di Fisica, Universit\`a degli Studi di Modena e 
Reggio Emilia, Modena, Italy}

\begin{abstract}
We study the crossover between liquid and solid electron phases
in a two-dimensional harmonic trap as the density is progressively
diluted. We infer the formation of geometrically ordered phases 
from charge distributions and pair correlation functions
obtained via a large scale configuration interaction calculation.
\end{abstract}

\end{frontmatter}

\section{Introduction}

Semiconductor quantum dots \cite{Jacak,ReimannRMP,RontaniNato}
(QDs) are nanometer sized region of
space where free carriers are confined by electrostatic fields.
Almost all QD-based applications rely on electronic correlation
effects which are prominent in these systems. The dominance of
interaction in QDs is evident from the
multitude of strongly correlated few-electron states measured or
predicted under different regimes: Fermi liquid, ``Wigner molecule''
(the precursor of Wigner crystal in two-dimensional bulk),
charge and spin density wave, incompressible state reminescent
of fractional quantum Hall effect in two dimensions 
\cite{ReimannRMP}.

The few-body  problem has been faced
mainly via Hartree-Fock method, Density Functional Theory,
Configuration Interaction (CI), and Quantum Monte Carlo (QMC)
\cite{Jacak,ReimannRMP}. Contrary to mean-field methods, 
CI and QMC methods allow for the treatment of Coulomb correlation with
arbitrary numerical precision, and, therefore, represent the
natural choice for strongly correlated regimes.
In addition, CI is a straightforward approach
with respect to QMC, and gives access to both ground and excited
states at the same time, its main limitation being its computational cost. 
Here we apply the CI method to the
the strongly correlated regime {\em
par excellence}, namely the crossover region between Fermi liquid
and Wigner crystal \cite{ReimannRMP}. We find evidence
of the formation of Wigner molecules, and confirm previous
QMC calculations \cite{Egger99,Reusch03}.

\section{The crossover between liquid and solid phases}

We consider a few electrons in a two-dimensional
harmonic trap,
which was proven to be an excellent model for different experimental
setups \cite{ReimannRMP}. The QD effective-mass Hamiltonian is:
\begin{equation}
H = \sum_i^N
\left[-\frac{\hbar^2}{2m^*}\nabla^2 + \frac{1}{2}m^*\omega_0^2\rho_i^2\right]
+\frac{1}{2}\sum_{i\neq j}\frac{e^2}
{\kappa|\boldsymbol{\rho}_i-\boldsymbol{\rho}_j|}.
\label{eq:HI}
\end{equation}
Here $m^*$ and $\kappa$ are the effective mass and the static relative 
dielectric constant of the host semiconductor, respectively, 
$\boldsymbol{\rho}_i\equiv (x_i,y_i)$ is the position of the
$i$-th electron, $\omega_0$ is the characteristic frequency of the trap. 
The cylindrical spatial symmetry group
of the system is $D_{\infty h}$, and the single-particle 
QD wavefunctions, the so called Fock-Darwin (FD) orbitals \cite{Jacak},  
have an azimuthal and a radial quantum number, respectively.
The good global quantum numbers of the few-body system are
the total angular momentum in the direction perpendicular to the plane, $M$,
the total spin, $S$, and its projection along the $z$-axis, $S_z$.

We solve numerically the few-body problem of Eq.~(\ref{eq:HI}), for 
the ground state at different numbers of electrons, $N$, by means of
the CI method \cite{Rontani04}:
we expand  the many-body wavefunction in a series of Slater determinants 
built by filling in the FD orbitals with $N$ electrons, and consistently 
with global symmetry constraints. On this basis, the Hamiltonian 
(\ref{eq:HI}) is a large, sparse matrix that we diagonalize by means of 
a Lanczos parallel routine \cite{ARPACK}.  
Note that, before diagonalization, we are able to build separate sectors 
of the Fock space 
corresponding to different values of $(M,S,S_z)$. To this aim, we rely on 
tabulated Clebsh-Gordon coefficients obtained, once for all, via numerical 
diagonalization of the $S^2$ matrix (cf.~Sec.~2.3 of Ref.~\cite{Pauncz79}).
We use, as a single-particle basis, up to 36 FD orbitals, and we need
to diagonalize matrices of linear dimensions up to $\approx 10^6$.
\begin{figure}[hmbt]
\centerline{\includegraphics[scale=0.4]{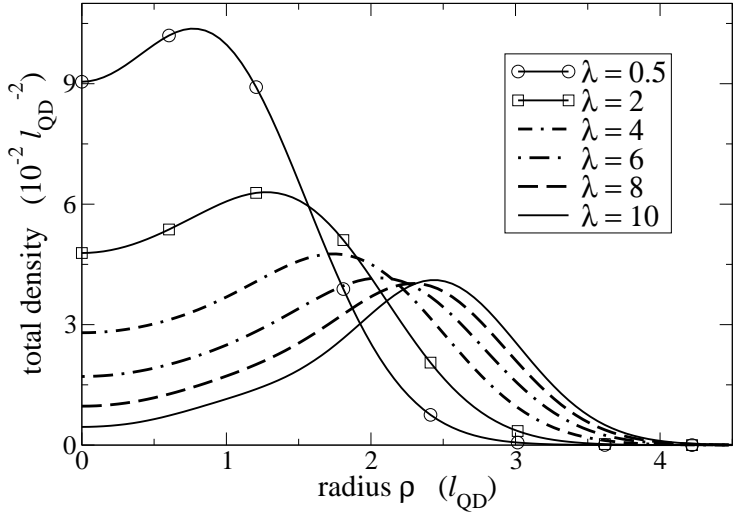}}
\caption{Normalized total density of the five-electron ground state
as a function of the radius, $\rho$, for different values of the
correlation strength parameter, $\lambda$. The length unit is the
characteristic dot radius, $\ell_{\mathrm{QD}}$.
}
\label{fig1}
\end{figure}

The way of tuning the strength of Coulomb
correlation in QDs is to dilute electron density.
At low enough densities, electrons pass from a ``liquid'' phase,
where low-energy motion is equally controlled by kinetic and
Coulomb energy, to a ``crystallized'' phase, reminescent of the
Wigner crystal in the bulk, where electrons are localized in space
and arrange themselves in a geometrically ordered configuration
such that electrostatic repulsion is minimized \cite{ReimannRMP,RontaniNato}.
The typical QD lateral extension is given by the characteristic
dot radius $\ell_{\mathrm{QD}} = (\hbar/m^*\omega_0)^{1/2}$,
$\ell_{\mathrm{QD}}$ being the mean square root of $\rho$ on the FD 
lowest-energy level.
As we keep $N$ fixed and increase $\ell_{\mathrm{QD}}$, the 
Coulomb-to-kinetic energy ratio 
$\lambda = \ell_{\mathrm{QD}}/a^*_{\mathrm{B}}$ [$a^*_{\mathrm{B}}
=\hbar^2\kappa/(m^*e^2)$
is the effective Bohr radius of the dot] \cite{Egger99} increases
as well, driving the system into the Wigner regime.

Figure \ref{fig1} shows the total electron density vs.~$\rho$, 
for $N=5$ and for different values of $\lambda$. The density is 
normalized to $N$, and the ground state remains 
$(M,S,S_z)=(1,1/2,1/2)$ across the liquid-solid transition. 
While in the non-interacting case ($\lambda\rightarrow 0$)
the charge is homogeneously spread across the dot, approximately 
between $\rho=0$ and $\rho=2\ell_{\mathrm{QD}}$, for $\lambda>4$ 
an outer ring develops while the weight in the inner region is 
stronlgy depleted. For $\lambda>8$ a well resolved
peak forms at $\rho\approx 2.5 \ell_{\mathrm{QD}}$, corresponding 
to the ``freezing'' of the electron liquid into a Wigner molecule
made of five electrons sitting at the corners of a regular pentagon.
This interpretation is supported by our analysis of the pair
correlation function \cite{RontaniNato} and by the excellent 
agreement between the computed values of ground-state energies 
of our calculation and those obtained
via QMC \cite{Egger99,Reusch03}.

\begin{figure}[t]
\centerline{\includegraphics[scale=0.4]{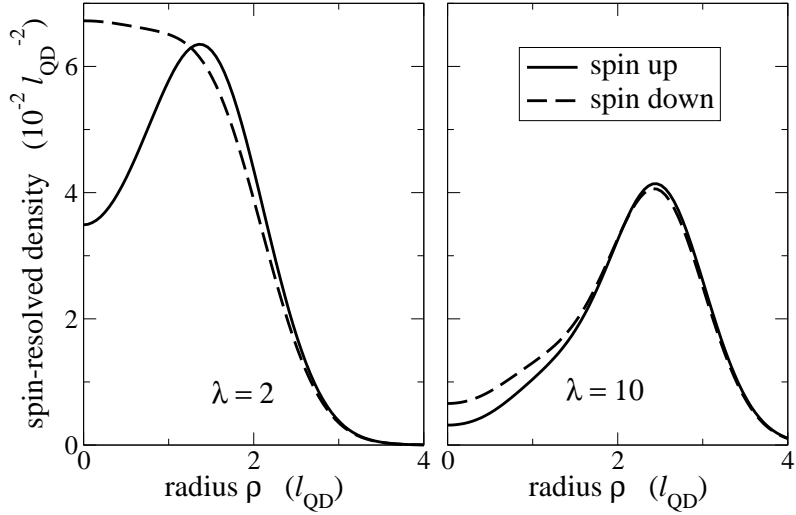}}
\caption{Normalized spin-resolved density of the five-electron 
ground state as a function of the radius, $\rho$. 
Left panel: $\lambda = 2$. Right panel: $\lambda = 10$. 
The length unit is the characteristic dot radius,
$\ell_{\mathrm{QD}}$.}
\label{fig2}
\end{figure}

The ``crystallization'' of the five-electron molecule is confirmed by the
analysis of the spin-resolved density, normalized to the number of electrons 
with a given spin [Fig.~\ref{fig2}].
In the liquid phase [left panel of Fig.~\ref{fig2}, $\lambda=2$], since 
the number of electron is odd, densities for the two spin species of course 
differ. In the Wigner limit (right panel, $\lambda=10$), instead, the system
becomes classical and the spin degree of freedom irrelevant: densites
for electrons with spin up or down coincide.

This paper is supported by MIUR-FIRB RBAU01ZEML, 
MIUR-COFIN 2003020984,
Iniziativa Trasversale INFM Calcolo Parallelo 2004, MAE-DGPCC.

\bibliography{rontani-ccp04}

\end{document}